\documentclass[a4paper,11pt]{article}
\pdfoutput=1
\usepackage{jheppub}
\usepackage[T1]{fontenc}

\newcommand{\be}{\begin{equation}}
\newcommand{\ee}{\end{equation}}
\newcommand{\ba}{\begin{eqnarray}}
\newcommand{\ea}{\end{eqnarray}}
\newcommand{\n}[1]{\label{#1}}
\newcommand{\eq}[1]{Eq.(\ref{#1})}

\newcommand{\hh}{\, ,\hspace{0.5cm}}
\newcommand{\hhh}{\, ,\hspace{0.2cm}}
\newcommand{\BM}[1]{{\mbox{\boldmath $#1$}}}
\newcommand{\ve}{\varepsilon}

\title{Information loss problem and a `black hole' model with a closed apparent horizon}
\author{Valeri P. Frolov}
\affiliation{Theoretical Physics Institute, University of Alberta,
Edmonton, AB, Canada,  T6G 2G7}
\emailAdd{vfrolov@ualberta.ca}

\abstract{In a classical description the spacetime curvature inside a black hole infinitely grows. In the domain where it reaches the Planckian value and exceeds it the Einstein equations should be modified. In the absence of reliable theory of quantum gravity it is instructive to consider simplified models. We assume that a spacetime curvature is limited by some value (of the order of the Planckian one). We use modified Vaidya metric, proposed by Hayward, to describe the black hole evaporation process. In such a spacetime the curvature near $r=0$ remains finite, it does not have an event horizon and its apparent horizon is closed. If the initial mass of such a `black hole' is much larger than the Planckian one its properties (as seen by an external observer) are practically the same as properties of the `standard' black hole with the event horizon. We study outgoing null rays in the vicinity of the outer apparent horizon and introduce a notion of a quasi-horizon. We demonstrate that particles, trapped inside a `black hole' during the evaporation process, finally may return to external space after the evaporation is completed. We also demonstrate that such quanta would have very large blue-shift. The absence of the event horizon makes it possible restoration of the unitarity in evaporating black holes.}

\begin{document}
\maketitle
\flushbottom

\section{Introduction}

Information loss puzzle is one of the longstanding problems of the black hole physics. A black hole can be formed by matter in a pure quantum state. In the subsequent process of its quantum evaporation its radiation is described by a (thermal) density matrix. If the black hole completely disappears what is left is its emitted radiation. Hence, in such a process the pure initial quantum mechanical state is transformed into a state described by the density matrix, so that unitarity would be violated. In order to describe such processes in quantum mechanics the latter should be modified. For this purpose Hawking in 1976 \cite{Hawk:76} introduced a so called superscattering operator. The information loss puzzle (the loss of the unitarity in processes involving black holes) was (and still is) the subject which has attracted a lot of attention (see e.g. \cite{Pres:92,Gidd:92,Page:93a,Page:93b,Strom:94,FrNo:98,Math:09,HoSm:10} and references therein).

The string theory strongly advocates the point of view that the unitarity should be restored after the evaporation of the black hole. These arguments are based on the AdS/CFT duality according to which the evolution of the black hole in the bulk asymptotically AdS spacetime is in correspondence with the unitary evolution of the corresponding quantum field on the boundary (see e.g. \cite{HoMa:04} and references therein). However attempts to find the mechanism of the unitarity restoration results in new problems. In order to solve the latter  it was recently proposed a `firewall' model \cite{AlMaPoSu:13}. According to this model any attempt by a falling into a black hole observer to determine an exact quantum state of the matter results in a creation of a firewall which changes the structure of the black hole itself. However this drastic approach seams to be not very appealing. Critical discussion of the firewall model can be found in \cite{Suss:12,Jaco:13}. Hawking in his recent paper \cite{Hawk:14} proposed to solve the information loss puzzle by assuming that in the gravitational collapse an apparent horizon is formed but the event horizon does not exists.

Such models have been studied earlier. A black hole model with a closed apparent horizon was proposed more than  30 years ago in \cite{FrVi:79,FrVi:81}. Later similar models have been widely discussed (see e.g. papers\cite{RoBe:83,Solo:99,Hayw:06,Anso:08,BaMaMo:13,LuSt:13,RoVi:14} and references therein). Such models also naturally arise in a study of the gravitational collapse in quantum gravity \cite{KaSo:93,Haji:02,Grum:03,Grum:04,ZiKu:10}. The purpose of the present paper is to review main features of the models with a closed apparent horizon. For this purpose we use a simple model, proposed by Hayward \cite{Hayw:06}, which allows one to study these properties in detail.

The paper is organized as follows. In section~2 we discuss our main assumptions. Sections~3--4 describe properties of the  model based on the modified Vaidya metric. In section~5 we introduce a notion of a {\em quasi-horizon}, which is a natural generalization of the event horizon for evaporating black holes. Namely, for slow decrease of the `black hole' mass it is `almost null'. It also serves as a separatrix between outgoing null rays that reach infinity and those that propagate to the center of the `black hole'. Behavior of null rays in the vicinity of the inner horizon is considered in section~6. Global properties of the `black hole' model with the closed apparent horizon are discussed in section~7. The last section~8 contains general discussion of the models with a closed apparent horizon in relation to the information loss puzzle. We also discuss some of the further problems of such models and their possible solutions. In the Appendix we demonstrate how a simple massive thin shell model can be used for an effective description of the quantum particle creation by a black hole.

\section{Assumptions}

We consider a formation of a spherical black hole in a gravitational collapse of a massive object and its subsequent quantum evaporation. We assume that the initial mass $M$ of the black hole is much larger than the Planckian mass $m_{Pl}$, so that the black hole is a classical object. We also assume that its gravitational field during all the evolution (including the final state of the evaporation) as well in the black hole interior is described by a metric tensor $g_{\mu\nu}$. In the domain where quantum corrections are small it obeys the Einstein equations. At the initial stage of the evaporation one can use quasi-classical description, so that the Hawking process results in the positive energy flux of created particles to infinity and (in accordance with the conservation law) by the negative energy flux through the horizon. The latter decreases the mass of the black hole. The main contribution to this process is due to massless fields, so that we describe the energy flux to future null infinity by a properly chosen null fluid stress-energy tensor. We assume that the negative energy flux through the black hole horizon into its interior can also be approximated by the null fluid with negative energy. Both incoming and outgoing fluxes are the result of the quantum process of particle creation in the vicinity of the horizon. To match these two fluxes and make the model consistent one should introduce a transition region between them. We assume that this region is narrow and use a model of a massive thin shell. We demonstrate in the Appendix that such a shell can be adjusted so that for slowly evolving black holes its effective stress-energy tensor is small.

In the spacetime domain, where the curvature becomes comparable with the Planckian one, the classical Einstein equations should be modified as a result of the effects of the vacuum polarization and intensive quantum particle creation. For the black hole with the gravitational radius $r_S=2M$ the curvature reaches the Planckian value at $r=r_0$ which can be found from the equation
\be
{r_S\over r_0^3}\sim l_{Pl}^{-2}\, .
\ee
In the spacetime domain where $r<r_0=r_s (l_{Pl}/r_s)^{2/3}$ the curvature calculated for the classical solution exceeds the Planckian one.

During the initial stage of the evaporation this region is deep inside the black hole. Modifications of the classical Einstein equations are also certainly required in order to describe the final stage of the evaporation. Namely in these domains the metric is expected to be quite different from its `classical form'. Since we do not have reliable theory of gravity valid at the Planckian scales we use special ansatz for the metric in these domains. We specify a model so that it satisfies the limiting curvature principle \cite{Markov:82,Markov:84}.
This prescription certainly contains an ambiguity. However even consideration of such simple models is quite instructive and allows one to study their robust predictions.

\section{A model}

The negative energy flux through the horizon, that accompanies the  particle creation, reduces the mass of the black hole. We use Vaidya solution to describe it
\be\n{vaidya}
dS^2=-F dV^2+2dV dr+r^2 d\omega^2\, ,\ F=1-{2M_-(V)\over r}\, .
\ee
The Ricci tensor for this metric is
\be
R_{\mu\nu}={2\dot{M}_-\over r^2}V_{,\mu}V_{,\nu}\, .
\ee
Since $V_{,\mu}$ is a null vector, one has $\BM{R}^2\equiv R_{\mu\nu}R^{\mu\nu}=0$ and $R=0$, while the square of the Riemann tensor is
\be
{\cal R}^2\equiv R_{\mu\nu\alpha\beta}R^{\mu\nu\alpha\beta}={48 M_-(V)^2\over r^6}\, .
\ee
Since we assumed that the curvature is limited by the Planckian value, the metric \eq{vaidya} should be modified at $r\sim r_0=l_{Pl} (M_-/l_{Pl})^{1/3}$.

To describe an evaporating black hole with a regular interior we use a modified Vaidya metric, proposed by Hayward \cite{Hayw:06}\footnote{The author is grateful to James Bardeen. After the first version of this paper appeared in archive, he attracted my attention to the paper by Sean Hayward, who has used the same form of the metric for the discussion of properties of `black holes' with a closed apparent horizon.}
This metric has the same form as \eq{vaidya} with a modified function $F$
\be\n{modvm}
F=1-{2M_-(V)r^2\over r^3+2M_-(V) b^2}\, .
\ee
Here $b$ is the cut-off parameter of the order of the Planck length. For $M_-\gg b$ and $r\gg r_0$ the modified metric remains practically the same as earlier. However for $r\ll r_0$
\be
F\sim 1-(r/b)^2\, .
\ee
This implies that the geometry near $r=0$ is regular and  $r=0$ is a regular timelike line. Denote
\be
q={2M_- b^2\over r^3}\, ,
\ee
then the curvature invariants for the modified Vaidya metric are of the form
\ba
&&{\cal R}^2={12 q^2(1-4q+18q^2-2q^3+2q^4)\over b^4 (1+q)^6}\,  ,\\
&&\BM{R}^2={18 q^4(5-2q+2q^2)\over  b^4 (1+q)^6}\,  ,\\
&&R={6q^2 (2q-1)\over b^2 (1+q)^3}\, .
\ea
For small $q$ one has
\be
{\cal R}^2\sim {12 q^2\over b^4}={48 M_-^2\over r^6}\hhh
\BM{R}^2\sim {18 q^4\over b^4}={288 M_-^4 b^4\over r^{12}}\hhh
R\sim -{6q^2\over b^2}=-{24 M_-^2 b^2\over r^6}\, .
\ee
In the limit $b\to 0$ one restores the expressions for the original Vaidya metric. In the opposite case (when $q\to \infty$) one has
\be
{\cal R}^2\sim 24 b^{-4}\hh \BM{R}^2\sim 36 b^{-4}\hh
R\sim 6 b^{-2}\, .
\ee
These relations show that the curvature of the modified Vaidya metric is limited and its maximal value is of the order of $b^{-2}$.

Before studying properties of the modified Vaidya metric it is convenient to present it in the dimensionless form. For this purpose we use the cut-off parameter $b$ as a natural length scale and introduce the following dimensionless quantities
\be
v={V\over b}\hh \rho={r\over b}\hh \mu(v)={M_-(V)\over b}\hh ds^2=b^{-2}dS^2\, .
\ee
The dimensionless form of the modified Vaidya metric is
\be\n{mvm}
ds^2=-f dv^2+2dv d\rho+\rho^2 d\omega^2\hh
f=1-{2\mu(v)\rho^2\over \rho^3+2\mu(v)}\, .
\ee

\section{Apparent horizon}

The apparent horizon for the metric \eq{mvm} is determined by the condition
\be
(\nabla \rho)^2\equiv f=0\, .
\ee
This gives the a following equation for the position of the apparent horizon
$\rho(v)$ at the moment of advanced time $v$ when the mass is $\mu(v)$. We write this relation in the form
\be\n{eqr}
2\mu(v)=Q(\rho)\hh Q(\rho)={\rho^3\over \rho^2 -1}\, .
\ee
The function $Q$ is positive for $\rho>1$, it has minimum at $\rho=\rho_*=\sqrt{3}$ and it grows infinitely when either $\rho\to 1$ or $\rho\to \infty$. This means that the apparent horizon does not exist for $\mu < \mu_*$ \footnote{It is interesting to notice that in the theory of gravity with quadratic in curvature corrections there exists a critical mass, so that an apparent horizon is not formed  if the mass a collapsing object is smaller than the critical one. This was demonstrated in \cite{FrVi:79,FrVi:81} for a model of  massive null shells. It is plausible that the existence of such a mass gap is a generic property of models with a closed apparent horizon. It would be interesting to study this problem for recently proposed class of singularity and ghost free theories of gravity \cite{BiGeKoMa:12}.
}, where
\be
\mu_*={1\over 2}Q(\rho_*)={3\sqrt{3}\over 4}\, .
\ee
(See also \cite{Hayw:06} and discussion therein)

For $\mu(v)>\mu_*$ the equation \eq{eqr} has two solutions $1<\rho_-(v)<\rho_+(v)$. We call the solution $\rho_-(v)$ an {\em inner branch} of the apparent horizon (or simply an inner horizon), and the solution $\rho_+(v)$ its {\em outer branch} (or an outer horizon). The inner horizon is located in the region $1<\rho_-(v)<\rho_*=\sqrt{3}$, that is in the spacetime domain where the curvature is of the order of the Planckian one. The outer horizon for large $\mu(v)$ is located close to $2\mu(v)$. The apparent horizon is closed (see Figure~\ref{Fig_2}). The region inside the closed horizon is called $T_-$-region.

\begin{figure}[tbp]
\centering
\includegraphics[width=8cm]{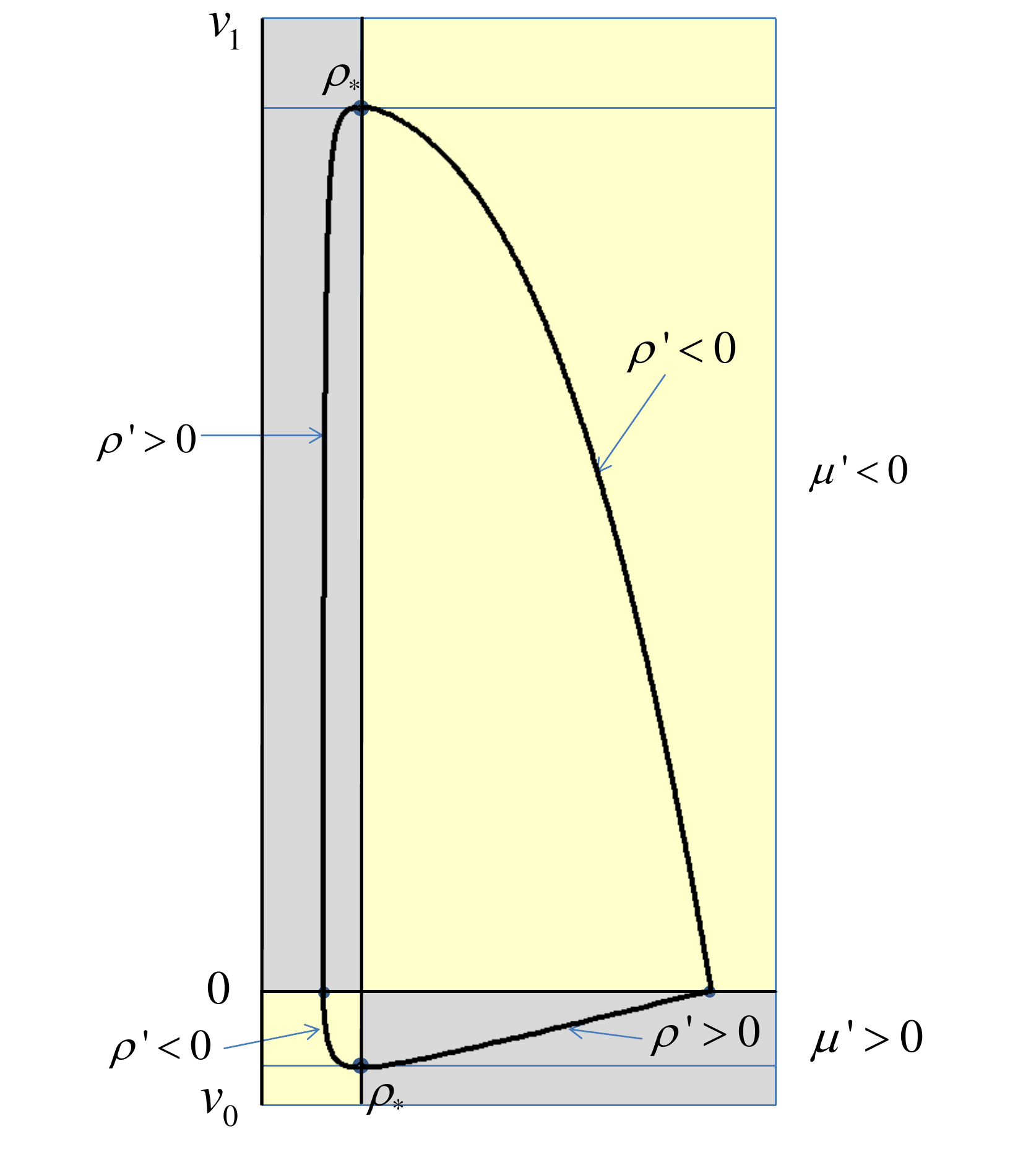}\hfill
  \caption{Apparent horizon structure.\label{Fig_1}}
\end{figure}

To specify the model one needs to choose the function $\mu(v)$. We shall do it later. At the moment we only assume that the `black hole' is formed as a result of the spherical collapse of the null fluid, and after its formation  its mass monotonically decreases as a result of the Hawking process and finally vanishes. For such a scenario the function $\mu(v)$ behaves as follows: it vanishes before $v_0$ (the moment when the collapse started), monotonically grows during the collapse and reaches its maximum value $\mu_0$ at $v=0$. After this it monotonically decreases until it vanishes at $v_1$ (the end of the evaporation). After this the mass $\mu(v)$ is identically zero. For such a scenario the apparent horizon is represented by a closed line in $(v,\rho)$ plane. It appears and disappears at the moments $v^*_{-}$ and $v^*_{+}$, respectively $(v_0<v^*_{-}<v^*_{+}<v_1)$. At the moments $v^*_{\pm}$ the inner and outer branches of the apparent horizon coincide
\be
\rho_{\pm}(v^*_{\pm})=\rho_*\, .
\ee
Using \eq{eqr} in the vicinity of this points one finds
\be
\rho\sim\rho_* +\lambda |v-v^*_{\pm}|^{1/2}\, ,\
\lambda=(4/3)^{3/4}|\mu_*'|^{1/2}\, .
\ee
For large $\mu$ one can solve \eq{eqr} perturbatively and get
\ba
\rho_+&=& {2\mu}-{1\over 2\mu}-{1\over 4\mu^3}+\ldots\, ,\n{rp}\\
\rho_-&=&1+{1\over 4 \mu}+{5\over 32 \mu^2}+{1\over 8\mu^3}+\ldots\, .\n{rm}
\ea

To describe Hawking evaporation of the black hole we use the following approximation
\be\n{mu}
\mu^3(v)=\mu_0^3-v\, ,
\ee
so that the rate of the mass loss is
\be\n{evap}
{d\mu\over dv}=-{1\over 3\mu^2}\, .
\ee
Restoring dimensionality one obtains
\be
{dM\over dV}=-C{m_{Pl}\over t_{Pl}}{m_{Pl}^2\over M^2}\, ,
\ee
with $C=1/3$. For a realistic black hole the coefficient $C$ depends on the number and statistics of the particles that are emitted. In our model \eq{mu} we neglect these details. For this choice of $\mu(v)$ the model contains two parameters: maximal mass of the `black hole' $\mu_0$ and time $v_0$ of its formation. For large mass $\mu(v)$ the rate of mass loss \eq{evap} is very small and the evaporation process is adiabatic with very high accuracy. Figure~1 demonstrates the structure of the apparent horizon for the model. Figure~2 shows the geometry of a flow generated by outgoing null rays in the modified Vaidya metric.

\begin{figure}[tbp]
\centering
\includegraphics[width=8cm]{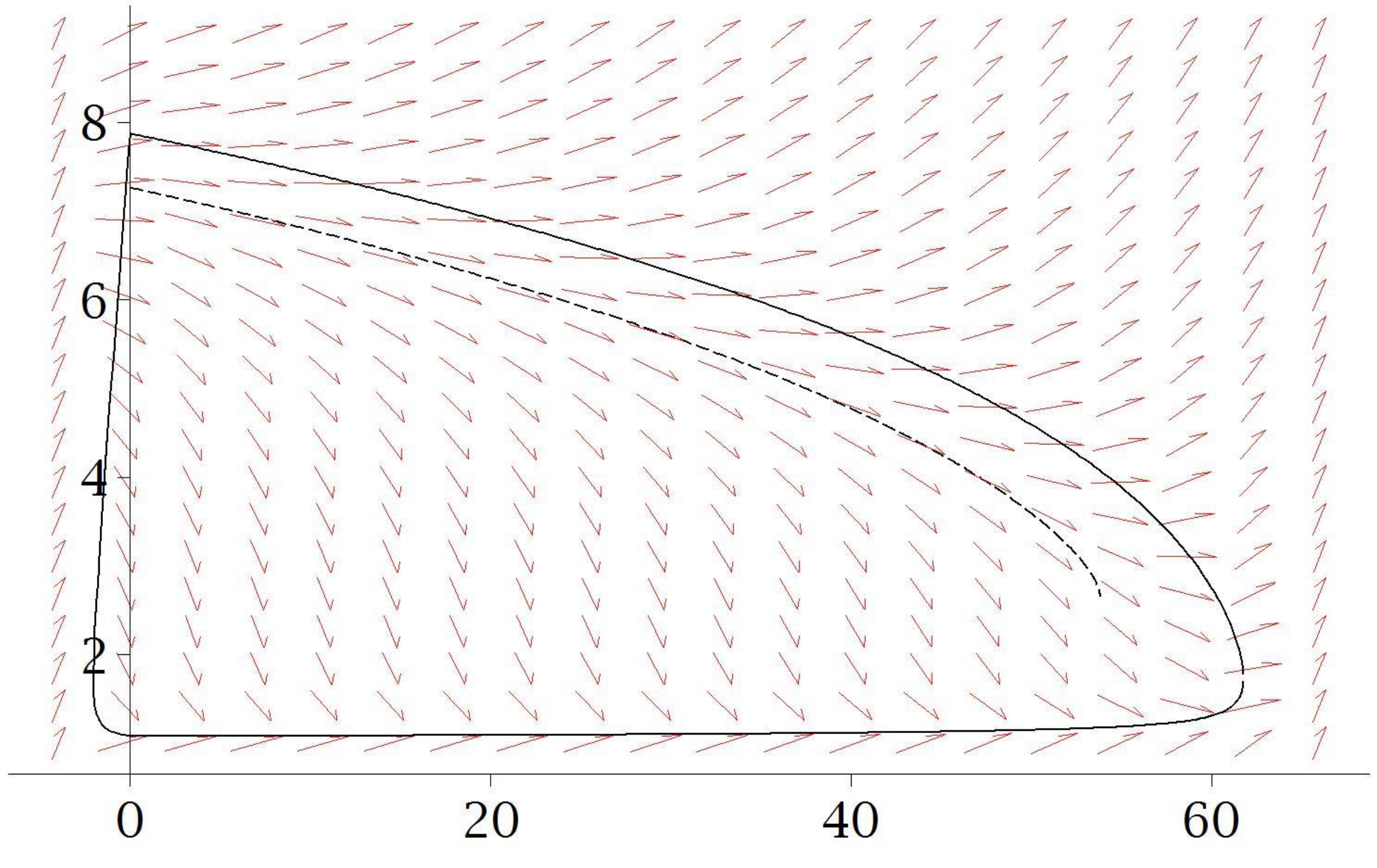}
  \caption{`Outgoing' radial null rays in the modified Vaidya metric. The horizontal axis is $v$ and the vertical one is $\rho$. The vector field $l^{\mu}=(1,f/2,0,0)$ tangent to the outgoing null rays is shown by small arrows. The plot is constructed for $\mu_0=4$ and $v_0=-3$. A closed solid line shows a position of the apparent horizon. A dashed line is a quasi-horizon.\label{Fig_2}}
\end{figure}

An external observer is not able to get information from the interior of such an object during all the time of its evaporation. However after the evaporation ends this becomes possible. This means that the event horizon does not exist in our model. According to the standard definition in the absence of the event horizon there is no black hole. This standard definition is well adapted for the proof of many important results about black holes, but it is based on the assumption that one can have information about future evolution of the matter in the Universe until  future time infinity. On the other hand the apparent horizon is defined for any given moment of time. Its existence indicates that the gravitational field is so strong that even light inside the apparent horizon moves to the `center'. Namely this property (very strong gravitational field) is important for explanation of the observable properties of black holes, that can be registered by a distant observer. However in the case of evaporating black holes the apparent horizon is a timelike surface. This means that it can be penetrated by light and particles in both directions, so that it looses the property of the boundary of the no-escape region.

In the next section we define a quasi-horizon which, for slowly evolving black holes, has the property of the boundary of `no-return' domain and which is practically a null surface. These features make quasi-horizon similar to the event horizon, but its definition does not require knowledge about infinite future. In order to make it clear that the objects with a closed apparent horizon do not have the event horizon, we use quotation marks and write a `black hole'.

\section{Quasi-horizon}

Consider an outgoing null ray passing through the apparent horizon at time $v_{\circ}$ when its size is $\rho_{\circ}=\rho(v_{\circ})$. If the mass of the black hole decreases with time, then for a slightly later time $v>v_{\circ}$ the gravitational field at $\rho_{\circ}$ becomes weaker. Hence such a ray propagates with an increase of the radius $r$ and finally it escapes to infinity. In order to characterize the boundary of the no-escape region we introduce a notion of a {\em quasi-horizon}. For black holes with changing mass this is a natural generalization of the notion of the event horizon. The first order differential equation for outgoing null rays is of the form
\be\n{onr}
{d\rho\over dv}={1\over 2}f(\rho,v)\, .
\ee
We define the quasi-horizon by the condition that $d^2\rho/dv^2$ for such a ray vanishes. This gives the following equation\footnote{It is easy to check that this condition implies that the gradient $\nabla_{\mu} f$ of the function $f$ is null at the quasi-horizon. Really,
\be
(\nabla f)^2=\partial_{\rho}f(2\partial_v f+f\ \partial_{\rho}f)=0\, .
\ee
}
\be\n{qh}
2\partial_v f+f\ \partial_{\rho}f=0\, .
\ee
Quasi-horizon for an evaporating black hole is shown at Figure~\ref{Fig_2}.
For a static metric $\partial_v f=0$ and the quasi-horizon coincides with the apparent and the event horizons, $f=0$.

For large $\mu(v)$ one has
\be
f=1-{2\mu(v)\over \rho}\, ,
\ee
and equation (\ref{qh}) for $\rho>0$ has a solution
\be
\rho=\rho_{qh}={4\mu(v)\over 1+\sqrt{1-16\mu'(v)}}\, .
\ee
Let us assume that $|\mu''|\ll |\mu'|$,  \footnote{This condition certainly is valid for $\mu\gg 1$. Really, $\mu'\sim \mu^{-2}$ and $\mu''\sim \mu^{-5}$ and $|\mu''|/\mu'^2\sim \mu^{-1}\ll 1$.}
then
\ba
&&{d\rho_{qh}\over dv}\approx {4\mu'\over 1+\sqrt{1-16\mu'(v)}}\, ,\\
&&f(\rho_{qh},v)={1\over 2}(1-\sqrt{1-16\mu'(v)})
\, .
\ea
It is easy to check that the relation \eq{onr} is valid for $\rho_{qh}(v)$ up to the terms of the order of $|\mu''|$. Hence for large $\mu(v)$ the surface $\rho=\rho_{qh}$ is practically null.

Let us discuss the properties of null rays propagating close to the quasi-horizon. We consider the `black hole' at time $v$ when its mass is much larger than the Planckian one, $\mu(v)\gg 1$. The radius of outer horizon is close to $2\mu(v)$ (see \eq{rp}), so that one can neglect the quantity $2\mu(v)/\rho^3$ and write
\be
f=1-{2\mu(v)\over \rho}\, .
\ee
Let us fix the moment of the advanced time $v_{\circ}$ and denote
\ba
&&\mu(v_{\circ})=\mu_{\circ}\hh (d\mu(v)/dv)|_{v_{\circ}}=\mu'_{\circ}\, ,\\
&&v-v_{\circ}=2\mu_{\circ} x\hhh \rho=2\mu_{\circ}y \hhh
z=-[x+(2\mu_{\circ}')^{-1}]
\, .
\ea
In these notations and for the chosen approximation the equation \eq{onr} for the outgoing null rays takes the form
\be\n{lineq}
{dy\over dz}=-{1\over 2}-{\mu'_{\circ} z\over y}\, .
\ee

The equation (\ref{lineq}) for null rays near slowly evolving apparent horizon was studied in \cite{VoZaFr:76}. It allows the following solution
\ba
&-&\ln(z/z_{\circ})=B(q)-B(q_{\circ})\hh q=y/z\, ,\n{eqBBQ}\\
B(q)&=&{q_+\over q_+-q_-}\ln |q-q_+|-{q_-\over q_+-q_-}\ln |q-q_-|\, .\n{BBQ}
\ea
Here
\be
q_{\pm}=-{1\over 4}\left( 1\pm \sqrt{1-16\mu'_{\circ}}\right)\, .
\ee
For $\mu'_{\circ}<0$ one has $q_+<0$ and $q_->0$.
Besides this general solution, there exists a special solution corresponding to the degenerate case $q=q_{-}=$const. This solutions is
\be
y=-q_{-}\left[x+(2\mu'_{\circ})^{-1}\right]\, .
\ee
It is easy to check that it coincides with the quasi-horizon. For $q\to q_-$ the function $B(q)$ becomes infinitely negative. Thus for outgoing null rays near the quasi-horizon one can approximate $B(q)$ as follows
\be
B(q)-B(q_{\circ})\approx -{q_-\over q_+-q_-}\ln \left|{q-q_-\over q_{\circ}-q_-}\right|\, .
\ee
In this approximation a solution of \eq{eqBBQ} is
\be\n{sol}
y=(1+2\mu'_{\circ}x)\left[ \hat{y} +(y_{\circ}-\hat{y})(1+2\mu'_{\circ}x)^{\beta}\right]\, .
\ee
Here $\hat{y}=-q_-/(2\mu'_{\circ})=-(2q_+)^{-1}$ and $\beta={q_+-q_-\over q_-}$ is a negative number.

If the initial value of $y$ coincides with the critical value, $y_{\circ}=\hat{y}$, the solution takes the form $y=q_- z$, that is it coincides with the quasi-horizon. If $y_{\circ}>\hat{y}$ it deviates from the critical solution to the right and the outgoing null ray propagates to infinity, while in the opposite case, $y_{\circ}<\hat{y}$, the ray  moves to the center. The parameter $\beta$ controls the rate of expansion and contraction of the rays near the quasi-horizon. If one traces outgoing null rays backward in time, one finds that  for such rays near a slowly evolving black holes the quasi-horizon plays the role of the attractor.

\begin{figure}[tbp]
\centering
  \includegraphics[width=9cm]{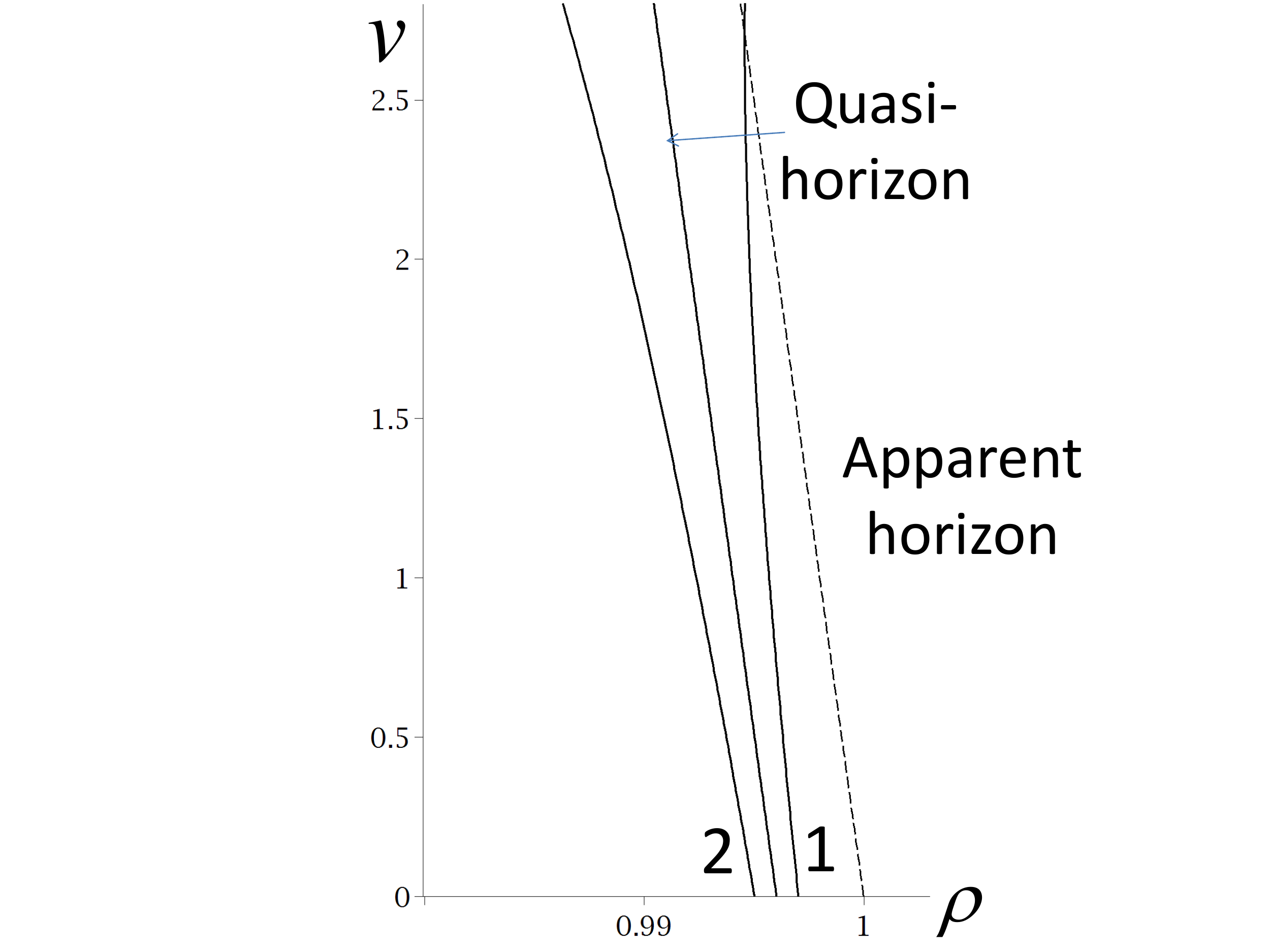}
  \caption{The quasi-horizon and the apparent horizon for a slowly evolving `black hole'. An outgoing photon 1 emitted outside the quasi-horizon propagates to infinity. A similar photon 2 emitted inside the quasi-horizon remains in the $T_-$-region during the time of the `black hole' evaporation.\label{Fig_3}}
\end{figure}

For small $\mu'_{\circ}$ one has
\ba
q_+&=& -{1\over 2}+\ldots\hhh
q_-= -2\mu'_{\circ}+\ldots\, ,\\
\beta&=&{1\over 4\mu'_{\circ}}+\ldots\hhh
\hat{y}=1+\ldots\, ,
\ea
where $\ldots$ denote higher order in $\mu'_{\circ}$ terms.
The relation \eq{sol} takes the form
\be
y-1\approx (y_{\circ}-1)(1+2\mu'_{\circ}x)^{\beta}\, .
\ee
In the limit $\mu'_{\circ}\to 0$ it can be rewritten as
\be
\ln{{y-1\over y_{\circ}-1}}=\lim_{\mu'_{\circ}\to 0}(2\mu'_{\circ} \beta) x=x/2\, .
\ee
Thus for a static black hole of mass $M$ one has
\be\n{stat}
r-2M=(r_{\circ}-2M)\exp[\kappa (V-V_{\circ})]\, ,
\ee
where $\kappa=(4M)^{-1}$ is the surface gravity. This relation correctly reproduces to a well known result for a static black hole.

\section{Inner horizon}

Let us discuss properties of the metric \eq{mvm} near the inner horizon. At first let us notice that in the limit $\mu\gg 1$ this metric only slightly differs from the de Sitter metric
\be
ds^2=-(1-\rho^2)dv^2 +2dv d\rho+\rho^2 d\omega^2\, ,
\ee
However, in the exact de Sitter metric outgoing null rays never (in time $v$) cross the horizon $\rho=1$. Important difference of the metric \eq{mvm}  in the vicinity of $\rho=1$ is that it is time dependent, and after the evaporation of the black hole the inner horizon disappears.

As earlier we denote by $\rho_-(v)$ the solution of \eq{eqr} obeying conditions $1<\rho_-<\rho_*=\sqrt{3}$. Let us fix a moment of advanced time $v_{\circ}$ and denote
\be
\mu(v_{\circ})=\mu_{\circ}\hhh\mu'(v_{\circ})=\mu'_{\circ}\hhh
v-v_{\circ}=2\mu_{\circ}x\hhh \mu\approx \mu_{\circ}(1+2\mu'_{\circ}x)\hhh \rho=\rho_{_-{\circ}}+y(x)\, .
\ee
Here $\rho_{_-{\circ}}$ is the value of the radius of the inner horizon at time $v_{\circ}$.
Substituting these relations into the equation for outgoing rays \eq{onr} and keeping the linear in $x$ and $y$ leading terms one obtains the following equation
\be
{dy\over dx}=Ay+Bx\, ,\
A=-{4\mu_{\circ}-3\rho_{\circ}\over 2\rho_{\circ}}\, ,\
B=-{\mu'_{\circ}\rho_{\circ}\over \mu_{\circ}}\, .
\ee
A solution of this equation is
\be\n{yyyy}
y=-{B\over A^2}-{B\over A}x+C e^{Ax}\, .
\ee
Since $\rho_{\circ}\approx 1\ll \mu_{\circ}$ one can slightly simplify expression \eq{yyyy} and write it in the form
\ba
&&y\approx \rho_{qh}^-+C e^{-(v-v_{\circ})}\, ,\n{yyy}\\
&&\rho_{qh}^-=\rho^-_{\circ}+{\mu'_{\circ}\over 4\mu_{\circ}^3}-{\mu'_{\circ}\over 2\mu_{\circ}}x\, .\n{inqh}
\ea
One can check that \eq{inqh} is the equation for the inner quasi-horizon. Relation \eq{yyy} implies that $\rho_{qh}^-$ is an attractor for outgoing null rays propagating in its vicinity.

\section{Global properties of the model}

\subsection{Null rays}

\begin{figure}[tbp]
\centering
  \includegraphics[width=9cm]{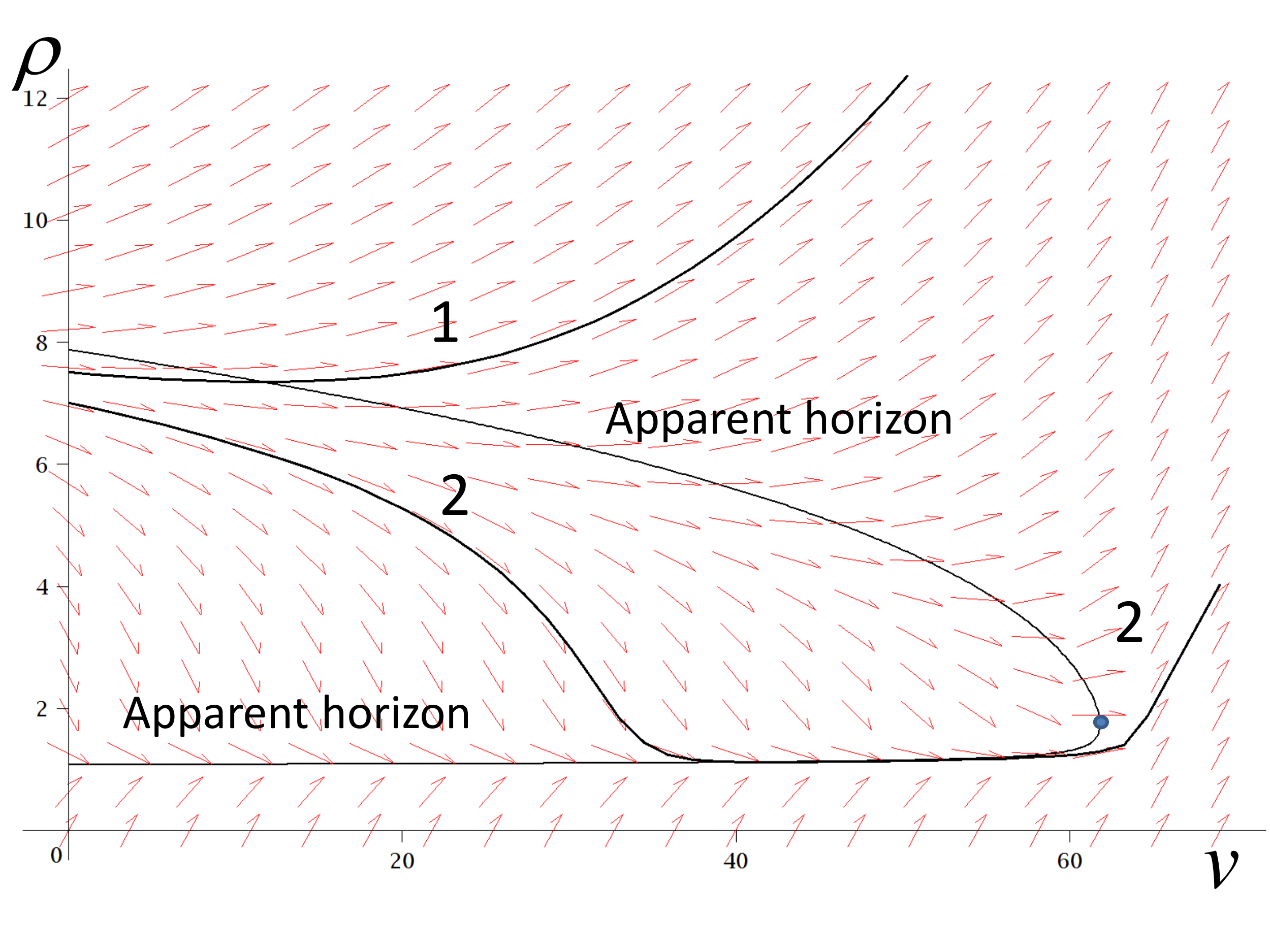}\\
  \caption{Outgoing null rays. An outgoing null ray 1 moves outside of the quasi-horizon. At later time it leaves the vicinity of the black hole and propagates to infinity. A far distant observer registers that it happens at some moment of the retarded time $u$. In the WKB approximation such a ray describes a trajectory of a `quantum'  emitted in the process of the Hawking radiation. The temperature of the thermal radiation registered by a distant observer at time $u$ is $(8\pi \mu(u))^{-1}$ \cite{VoZaFr:76}. A ray 2 represents a `partner' of the emitted Hawking quantum. It moves in $T_-$ region in the direction of increasing of $\rho$. After it reaches the inner horizon, it propages in its close vicinity until a complete evaporation of the black hole. \label{Fig_4}}
\end{figure}

\begin{figure}[tbp]
\centering
  \includegraphics[width=9cm]{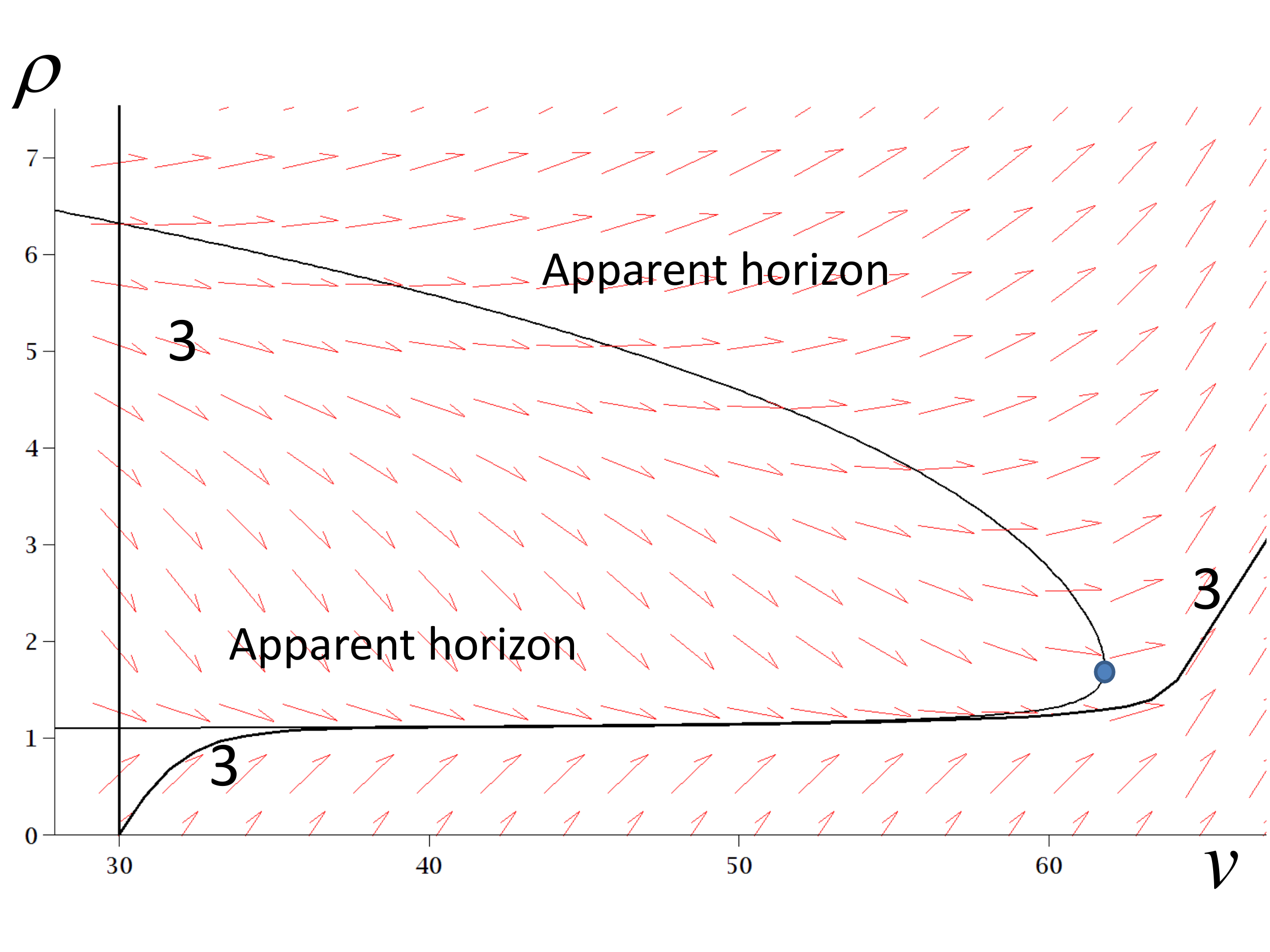}\\
  \caption{An incoming null ray 3 (at $v=30$) passes through $\rho=0$ and then propagates as an outgoing ray. It moves along the inner horizon until a complete evaporation of the black hole. \label{Fig_5}}
\end{figure}

Let us discuss general properties of the proposed black hole model with a closed apparent horizon. First of all, any particle that falls into such a `black hole' finally returns to external space after the black hole's complete evaporation. This means that in such a model there is no event horizon. However, if the initial mass is much larger than the Planckian one the properties of such an object are practically identical to properties of an evaporating black hole during long period of time of its evaporation.The quasi-horizon effectively plays the role of the event horizon. Any particle or photon that crosses the quasi-horizon inevitably moves to the center until it reaches a region with the Planckian curvature. In this region it propagates exponentially close to the inner quasi-horizon, and remains there until the time when the mass of the `black hole' reaches the Planckian value. After this it propagates back to an external observer. Figures~\ref{Fig_4} and \ref{Fig_5} illustrate global properties of different types of null rays in the metric (\ref{mvm}).

A modified Vaidya metric with $\mu(v)>\mu_*$  has trapped surfaces. According to Penrose theorem a spacetime with a trapped surface, which contains a non-compact Cauchy surface, must be geodesically incomplete, provided the weak energy condition is satisfied. Since our spacetime is regular and geodesically complete, the weak energy condition should be violated in the `black hole' interior. It is easy to check that this is really so. Namely, for outgoing null rays with $l^{\mu}=(1,f/2,0,0)$ the quantity
\be
R_{\mu\nu}l^{\mu}l^{\nu}={2\mu' \rho^4\over (\rho^3+2\mu)^2}
\ee
is negative when $\mu'<0$.

\subsection{Carter-Penrose diagram}

In order to illustrate properties of the model we used spacetime diagrams in $(v,\rho)$ coordinates, which are well suited for the adopted metric. However, the global causal structure of the spacetime is more transparent when it is presented in the form of the Carter-Penrose diagram. To construct such a diagram one needs first to rewrite the metric in the double null coordinates $(u,v)$, and, after this, to make transformation that `brings' the infinity into the finite domain.

The function $u(v,\rho)$ can be found by solving the equation
\be\n{eeqq}
du=b(v,\rho)(d\rho-{1\over 2} f(v,\rho)dv)\, .
\ee
Here $b(v,\rho)$ is a factor which is required since the expression in the brackets is not a complete differential. For our choice of the metric \eq{eeqq} cannot be solved analytically. We describe now how to find the map $(v,\rho)\to (u,v)$ numerically.
Consider a point $(v_0,\rho_0)$ and let $\gamma$ be an out-going null ray passing through this point. Its equation is
\be\n{enr}
{d\rho(v)\over dv}={1\over 2} f(v,\rho)\, .
\ee
At late time $v$ $\mu(v)=0$ and $f=1$, so that a solution of \eq{enr} is a straight line $\rho=(u_0+v)/2$, where $u_0$ is a constant retarded time. Hence, to find parameters $(u_0,v_0)$ for a given point $(v_0,\rho_0)$ one can introduce a function $U(v)=v-2\rho(v)$ and integrate the equation
\be
{dU\over dv}=1-f(v,{1\over 2}(v-U(v)))\, ,
\ee
with the initial condition $U(v_0)=v_0-2\rho_0$. For large value of $v$, where $\mu(v)$ vanishes, the function $U(v)$ becomes constant and it coincides with the retarded time $u_0$. This numeric procedure determines a map of $(v,\rho)$ coordinates to the double null coordinates $(u,v)$. These coordinates are normalized at infinity by the conditions
\be
\xi_t^{\mu}u_{,\mu}=\xi_t^{\mu}v_{,\mu}=1\, ,
\ee
where $\xi_t^{\mu}$ is an asymptotic timelike Killing vector.

\begin{figure}[tbp]
\centering
  \includegraphics[width=12cm]{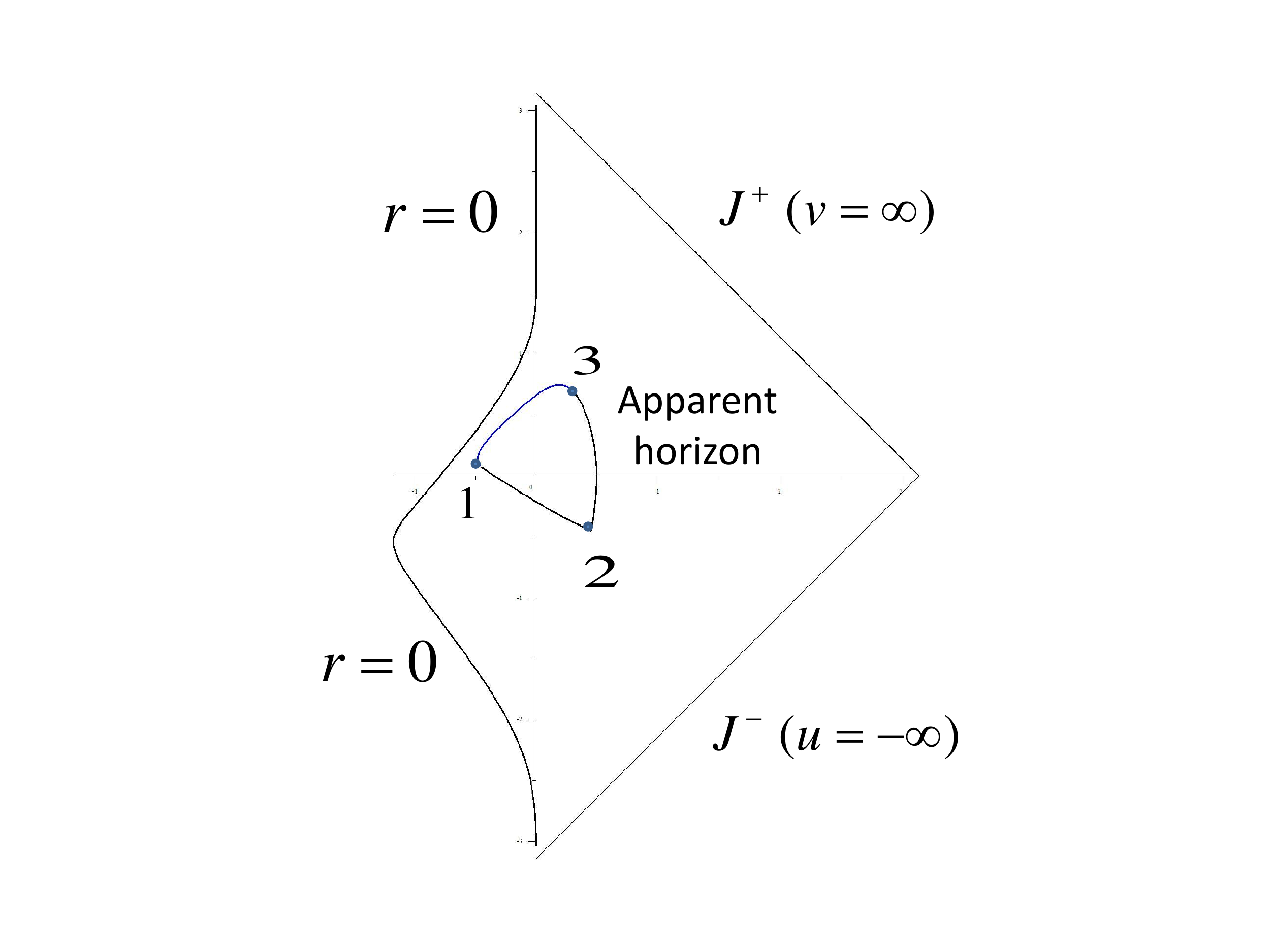}\\
  \caption{Conformal Carter-Penrose diagram for the spacetime of a non-singular `black hole' with a closed apparent horizon. Line 1-2 is the apparent horizon inside the collapsing matter. Lines 1-3 and 2-3 are the inner and outer horizons of the evaporating `black hole', respectively. This particular plot is constructed for the metric with the following parameters : $\mu_0=1.7$, $v_0=-4$. We also choose $q=5$. For larger mass $\mu_0$ the diagram is similar. \label{CP}}
\end{figure}

To compactify the coordinates $(u,v)$ we introduce new coordinates
\be\n{cpd}
\zeta=\arctan (v/q)-\arctan (u/q)\hh
\eta=\arctan (v/q)+\arctan (u/q)\, .
\ee
In these coordinates null lines $u=$const and $v=$const are represented by straight lines with the angle of their slope equal to $\pm \pi/2$. The parameter $q$ is an arbitrary positive constant which can used for better presentation of the diagrams. The future null infinity ${\cal J}^+$ in these coordinates is represented by a segment of the line $\zeta+\eta=\pi$, while the past null infinity is a segment of the line $\zeta-\eta=\pi$.
The conformal Carter-Penrose diagram for the modified Vaidya metric \eq{mvm} with the closed apparent horizon is shown a Figure~\ref{CP}.

\subsection{Frequency shift}

The inner horizon is an attractor for outgoing null rays in its vicinity. The exponential convergence of such rays to the inner horizon results in the exponential increase of the photon's frequency. This effect can be easily analyzed. Denote by $\alpha$ an affine parameter along such a ray. One of the geodesic equations in this affine parametrization gives
\be\n{ingr}
{d^2 v\over d\alpha^2}=-\kappa \left({dv\over d\alpha}\right)^2\hh
\kappa={1\over 2}\partial_{\rho}f \, .
\ee
\eq{ingr} can be written in the form
\be\n{eeqq}
{d\over dv}{dv\over d\alpha}=-\kappa {dv\over d\alpha}\, .
\ee
For our model the surface gravity $\kappa$ of the inner horizon with high accuracy is $-1$, so that a solution of \eq{eeqq} is
\be\n{freq}
{dv\over d\alpha}=\left.{dv\over d\alpha}\right|_{v_{\circ}}\exp(v-v_{\circ})\, .
\ee
To restore dimensionality in this relation we use a transformation
\be
V=b v\hh \lambda=b \alpha\, .
\ee
We  use also a scale ambiguity in the choice of the `new' affine parameter $\lambda$ and choose it so that
\be
\left.{dV\over d\lambda}\right|_{V_{\circ}}=\omega_{\circ}\, ,
\ee
where $\omega_{\circ}$ is the `physical' frequency of the `photon' at time $V_{\circ}$. Then \eq{freq} determines the frequency $\omega$ of a `photon' propagating in the vicinity of the inner horizon at later time $V$
\be
\omega=\omega_{\circ}\exp[(V-V_{\circ})/b]\, .
\ee
This confirms the above conclusion that the frequency of `photons' propagating close to the inner horizon becomes exponentially blue-shifted \cite{BoFr:86}.

\section{Discussion}

We presented a simple model of a `black hole' with a closed apparent horizon. We assume that the `black hole' is isolated and no matter falls into it after its formation. However the Hawking radiation reduces the mass. If the initial mass $M$ of the `black hole' is much larger than the Planckian mass, $M/m_{Pl}\gg 1$, this process is very slow. This adiabatic phase ends after the time of the order $\sim t_{Pl} (M/m_{Pl})^3$.

The model is constructed so that the curvature invariants in the interior of the `black hole' are limited. As a result the apparent horizon in the adopted model is closed and the metric in the `core' of the `black hole' is close to the de Sitter metric. This makes this model similar to the one proposed in \cite{FMM:89,FMM:90}. However their global properties are different. Instead of a new universe formation inside the black hole,  the black hole interior for the present model remains causally connected with the original external space.

The outer apparent horizon for an evaporating `black hole' is a timelike surface. Hence it looses the main property of the event horizon as a no-escape surface. We introduced a new notion -- the quasi-horizon,  and demonstrated that it is much better generalization of the event horizon for time-dependent `black holes' than the apparent horizon. The quasi-horizon separates two families of outgoing null rays: (1) The rays that enter $T_-$ domain and remain there until complete evaporation of the `black hole',  and (2) the rays that escape from the vicinity of the `black hole' and reach infinity. For adiabatic evolution of the `black hole' the quasi-horizon is practically a null surface. One can define a `black hole' as an object that has a quasi-horizon. It should be emphasized that the quasi-horizon is defined locally, and one should not wait infinite time and know all the details of the future evolution of the Universe, in order to make a decision whether one has a `black hole' or does not. This makes this new definition of the `black hole' more attractive than the `standard' one and more useful for the discussion of evaporating black holes.

Main feature of the presented model is that after the complete evaporation of the `black hole' all the information concerning the matter, that originally have produced it as well as particles created inside it in the Hawking evaporation process, becomes available to an external observer. This opens a possibility of the restoration of the unitarity in evaporating `black holes'.

One of the potential difficulty of the proposed model is strong blue-shift effect for rays propagating in the vicinity of the inner horizon  \cite{BoFr:86}. These rays are focused and produce exponentially thin beams of ultra-high frequency. Let us notice that this phenomenon has common features with  a well known transplanckian problem in black holes \cite{Jaco:91} and in the inflating universe \cite{MaBr:01}. Namely if one traces backward in time outgoing rays close to the event horizon of the black hole, one discovers a similar effect of focusing and exponentially large blue-shift. One can say that in models with a closed apparent horizon there exists  some kind of duality between properties of the inner and outer horizons. This duality
implies that one can expect that there exist an inverse Hawking process of annihilation of two quanta of positive and negative energy propagating on both sides of the inner horizon.

One also needs to remember that the proposed model is certainly over-simplified. We used a classical metric to describe the spacetime geometry. In quantum gravity such a metric is an effective one, that arises as a result of averaging of the quantum fluctuating metric and it obeys some (unknown at the moment) effective gravitational equations. Our assumption was that solutions of these modified Einstein equations obey the principle of the limiting curvature. But this description can be valid only when quantum fluctuations are small. The inner `core' of the `black hole' is of the Planckian size and, as a result of focusing, light rays propagate at much  smaller distance from the inner horizon. One can expect that quantum fluctuations of the metric are large at these scales and the description in terms of the effective metric may not be suitable.

Another important feature of the models with a closed apparent horizon is that the matter, that contains information about the state of the collapsing matter and phases of the Hawking quanta, is compressed in a small domain of the Planckian size and has very high (Planckian) density. This means that the non-linear quantum interaction between particles as well as their interaction with gravity becomes important. Certainly the presented model is too simplified to address these problems. Let us also mention that if the low energy gravity is an emergent phenomenon, then for proper description of the matter in the inner `core' and near the inner horizon one needs to use the language of the fundamental background theory of some heavy constituents (e.g strings).

All these problems are connected with unknown properties of the quantum gravity, the theory which does not exist at the moment. However a proposed model of a `black hole' with a closed apparent horizon allows one to address them and test some of possible new ideas.

\appendix

\section{Near horizon geometry}

As a result of the Hawking process a black hole emits radiation to infinity. Massless particles production dominates in this process. The outgoing energy flux can be effectively described by properly chosen null fluid. The conservation law requires that this radiation is accompanied by the negative energy flux through the horizon, which we also approximate by a null fluid. In order to make a model consistent one needs to assume that between the two regions with pure outgoing and pure incoming fluxes there exists a transition region, corresponding to the domain where the particle are created. We assume that this region is narrow and approximate it by a massive thin shell (see also discussion in \cite{Hayw:06}). In this appendix we demonstrate consistency of such a model and show that for slowly evolving black holes the parameters of the shell can be very small.

To describe the metric near a horizon we use the Vaidya solution which we write in the form
\be
dS_{\ve}^2=-F_{\ve}dZ_{\ve}^2 -2\ve dZ_{\ve} dr+r^2 d\omega^2\, ,\  F_{\ve}=1-{2M_{\ve}(Z_{\ve})\over r}\, .
\ee
Here $d\omega^2$ is the metric on a unit sphere and $\ve=\pm 1$. For $\ve=+1$ $Z_{+}$ is a retarded null time $U$, while for $\ve=-1$ $Z_{-}$ is an advanced null time $V$. The Ricci tensor calculated for this metric is
\be
R^{\ve}_{\mu\nu}=-{2\ve\over r^2}{dM_{\ve}\over dZ_{\ve}} Z_{\ve;\mu}Z_{\ve;\nu} \, .
\ee
If ${dM_{\ve}/dZ_{\ve}}<0$ then according to the Einstein equations  the corresponding stress-energy tensor for $\ve=+1$
\be
T^{+}_{\mu\nu}=-{1\over 4\pi r^2}{dM_+\over dU} U_{,\mu}U_{,\nu}
\ee
describes outgoing null fluid with positive energy density. The corresponding
stress-energy tensor for $\ve=-1$
\be
T^{-}_{\mu\nu}={1\over 4\pi r^2}{dM_-\over dV} V_{,\mu}V_{,\nu}
\ee
describes incoming null fluid with negative energy density.

Consider a spherical surface $\Gamma$ described by the equation
\be
Q(r,Z)=r-R(Z)=0\, ,
\ee
and assume that outside this surface the metric is $dS_+^2$ while inside it is $dS_-^2$. We also assume that the null coordinate $V$ in the inner region is chosen so that on $\Gamma$ it coincides with the corresponding value of $U$, so that one has
\be
V|_{\Gamma}=U|_{\Gamma}=Z\, .
\ee
The intrinsic 3-geometries induced by the metrics $dS_{\pm}^2$ on $\Gamma$ are
\be
d\sigma_{\ve}^2=- \left[1-{2M_{\ve}(Z)\over R}+2\ve R'\right] dZ^2+R^2 d\omega^2\, .
\ee
Here $R'=dR/dZ$. Both metrics are identical when the following condition is satisfied
\be
M_+(Z)-M_-(Z)=2RR'\, .
\ee
In what follows we assume that this condition is satisfied and write $M_{\pm}$ in the form
\be
M_{\ve}(Z)=M(Z)+\ve RR'\, ,
\ee
so that the induced metric on $\Gamma$ (which is the same for $\ve=\pm1$) is
\be
d\sigma^2=- (1-{2M(Z)\over R}) dZ^2+R^2 d\omega^2\, .
\ee

Denote by $n^{\ve}_{\mu}$ a unit normal vector to the surface $\Gamma$. Simple calculations give
\ba
n^{\ve}_{\mu}&=&\alpha^{-1} (-R',1,0,0)\, ,\\
\alpha_{\ve} &=&\sqrt{1-2M_{\ve}(Z_{\ve})/r+2\ve R'(Z_{\ve})}\, .
\ea
We denote by $e_{\hat{t}}^{\mu}$ a unit future directed vector tangent to $\Gamma$
\be
e_{\hat{t}}^{\mu}=\beta^{-1} (1,R',0,0)\, ,\ \ \beta=\alpha_{\ve}|_{\Gamma}=\sqrt{1-{2M(Z)\over R(z)}}
\, .
\ee
We also denote by $e_{\theta}^{\mu}$ and $e_{\phi}^{\mu}$ two other unit vectors tangent to $\Gamma$
\be
e_{\hat{\theta}}^{\mu}=(0,0,R^{-1},0)\hh
e_{\hat{\phi}}^{\mu}=(0,0,(R\sin\theta)^{-1},0)\, .
\ee
The tetrad components of the extrinsic curvature are
\be
K^{\ve}_{\hat{i}\hat{j}}=e_{\hat{i}}^{\mu}e_{\hat{j}}^{\nu}n^{\ve}_{\mu;\nu}\, .
\ee
We choose a special form of $\Gamma$ and put $R=2M(Z)(1+w)$, where $w$ is a dimensionless  positive small parameter. Then calculations give the following expressions for the non-vanishing components of $K^{\ve}_{\hat{i}\hat{j}}$
\be
K^{\ve}_{\hat{t}\hat{t}}=A+\ve B\, ,\
K^{\ve}_{\hat{\theta}\hat{\theta}}=-2wA+\ve B\, ,\
K^{\ve}_{\hat{\phi}\hat{\phi}}=-2wA+\ve B\, ,
\ee
where
\be
A={1\over 4M\sqrt{w(1+w)^3}}\hh B={M'\sqrt{1+w}\over \sqrt{w} M}\, .
\ee

The jumps of the extrinsic curvature at $\Gamma$ determine parameters (mass and pressure) of the shell. These jumps are
\be
[K_{\hat{t}\hat{t}}]=[K_{\hat{\theta}\hat{\theta}}]=[K_{\hat{\phi}\hat{\phi}}]=2B\, .
\ee
One can interpret the corresponding distribution of matter as being connected with a region of particle creation. For a slow change of the black hole mass, that is when $|M'|\ll 1$, the influence of this matter on the black hole geometry is proportional $|M'|$ and hence is extremely small and can be neglected.

\acknowledgments

The author is grateful to the Natural Sciences and Engineering
Research Council of Canada and to the Killam Trust for their financial support.

\end{document}